\documentclass[conference]{IEEEtran}
\IEEEoverridecommandlockouts
        \usepackage[pdftex]{graphicx}
        \usepackage{epstopdf}
\usepackage{graphicx}
\usepackage{dcolumn}
\usepackage{bm}
\usepackage{latexsym}
\usepackage{amsmath,amssymb,enumerate,color,url}
\usepackage{amsthm}
\usepackage{ascmac}
\usepackage{url}
\usepackage{algorithmic}
\usepackage{array}
\usepackage[tight,footnotesize]{subfigure}
\usepackage[font=footnotesize]{subfig}
\usepackage{color}
\usepackage{url}

\newcommand{\dd}{\mathrm{d}}
 
 \newcommand{\ii}{\mathrm{i}}

\theoremstyle{definition}
\newtheorem{theorem}{Proposition} 

 \def\hat#1{\widehat#1}

\begin{document}

\title{On the Fundamental Equation of User Dynamics \\
and the Structure of Online Social Networks}

\author{
\IEEEauthorblockN{Masaki Aida}
\IEEEauthorblockA{
\textit{Tokyo Metropolitan University}\\
Hino 191-0065, Japan \\
aida@tmu.ac.jp}

\and
\IEEEauthorblockN{Chisa Takano}
\IEEEauthorblockA{
\textit{Hiroshima City University}\\
Hiroshima 731-3194, Japan \\
takano@hiroshima-cu.ac.jp}

\and
\IEEEauthorblockN{Masaki Ogura}
\IEEEauthorblockA{
\textit{Osaka University}\\
Suita Osaka 565-0871, Japan \\
m-ogura@ist.osaka-u.ac.jp }
}

\maketitle              

\begin{abstract}
Online social networks suffer from explosive user dynamics such as flaming that can seriously affect social activities in the real world because the dynamics have growth rates that can overwhelm our rational decision making faculties. 
Therefore, a deeper understanding of user dynamics in online social networks is a fundamental problem in computer and information science. 
One of the effective user dynamics models is the networked oscillation model; it uses a second-order differential equation with Laplacian matrix. 
Although our previous study indicates that the oscillation model provides us with a minimal but effective model of user interactions, there still remains the open problem as to the existence of a first-order fundamental differential equation that respects the structure of the original network. 
This paper fills in this gap and shows that, by doubling the dimension of the state space, we can explicitly but naturally construct a fundamental equation that fully respects the structure of the original network.
%
\end{abstract}
%
\section{Introduction}
The widespread adoption of information networks, has dramatically activated the exchange of information among individuals, 
and the dynamics of users in online social networks is beginning to have a major impact beyond online communities; 
social activities in the real world are being influenced. 
In particular, explosive user dynamics such as the flaming phenomenon that occurs in online social networks can spread 
far faster than human rational decision making can respond, which can cause major social unrest.
Therefore, understanding the dynamics of online social networks in an engineering framework is an urgent and important issue.

Studies into the dynamics of online social networks have examined various models that reflect the diversity of characteristics of user dynamics. 
User dynamics that describe the adoption and abandonment of a particular social networking service (SNS) have been modeled by the SIR model, 
which is a traditional epidemiological model, and the irSIR model, which is an extension of that model~\cite{Nekovee2007,Cannarella2014EpidemiologicalMO}.
The consensus problem including user opinion formation is typical of the dynamics in online social networks~\cite{Olfati-Saber2004,Wang2010}. 
This is modeled by a first-order differential equation with respect to time using a Laplacian matrix that represents the social network structure.
The differential equation used in this model is a sort of continuous-time Markov chain on the network.
First-order differential equations with respect to time are also used in modeling of the temporal change of social network structure (how to link or to delink the nodes), and there are models that change in a continuous-time Markov chain~\cite{Snijders2010}. 
In addition to theoretical modeling, user dynamics analysis based on real network observations has also been studied~\cite{Cha2009,Zhao2012}. 

This paper focuses on explosive user dynamics such as flaming, which is defined as the divergence of the intensity of user dynamics. 
Since both epidemiological models and continuous-time Markov chains on networks describe the transition to the final state (steady state), they cannot describe the divergence of the intensity. 
Moreover, it is difficult to clarify the structure of the theoretical model behind user dynamics from just an analysis of actual data.
User dynamics in online social networks, including the explosive user dynamics, is generated by interactions between users. 
It is difficult to fully understand the details of interactions between users, but we can apply the concept of the minimal model; it models the simple interactions exhibited by a wide type of user interactions. 
Based on the minimal model, it has been proposed to apply the oscillation model on networks to describe user dynamics in online social networks \cite{aida2018}.
In the oscillation model approach, network dynamics is described by the wave equation on networks. 
The oscillation energy of each node calculated from the oscillation model gives a generalization of node centrality and includes the 
conventional node centrality measures (degree centrality and betweenness centrality) \cite{wasserman,carrington,mislove} 
commonly used in network analysis \cite{takano2018}. 
Also, by considering that the occurrence of explosive user dynamics 
such as flaming in online social networks demonstrate the characteristic that the oscillation energy diverges with time, we can discuss appearance factors of explosive user dynamics in relation to the structure of the online social network \cite{aida2018}.

This paper examines a fundamental equation \cite{aida2018} of the oscillation model on networks. 
The fundamental equation can explicitly describe the causal relation of the influence of the network structure on user dynamics. 
We give solutions to two major unresolved issues with the fundamental equation. 
Specifically, we derive all solutions of the wave equation from the fundamental equation and draw a concurrence between the link structure of the networks (represented by the fundamental equation) and that of the wave equation. 
Surprisingly, they are solved naturally and simultaneously.

\section{Oscillation Model on Networks}
This section briefly summarizes the oscillation model on networks according to \cite{aida2018}.  

Let $G(V,E)$ be a simple directed graph (without self-loop and duplicated links) with $n$ nodes representing a social network, 
$V=\{1,\,\dots,\,n\}$ denote the set of nodes, and $E$ denote the set of links. 
Hereafter, node IDs are denoted by Roman characters $1 \le i,\,j \le n$ and the oscillation modes are denoted by Greek characters $0 \le \mu,\,\nu \le n-1$. 

For a pair of adjacent nodes $i,\,j\in V$, we let the link weight of directed link  $(i \rightarrow j) \in E$ be denoted by $w_{ij}$, 
and define the adjacency matrix $\bm{\mathcal{A}} = [\mathcal{A}_{ij}]_{1\le i,j \le n}$ as 
\begin{align}
\mathcal{A}_{ij} := \left\{
\begin{array}{cl}
w_{ij},&  \quad (i\rightarrow j) \in E,\\
0,& \quad (i\rightarrow j) \not\in E.
\end{array}
\right. 
\end{align}
In addition, if the nodal degree of out-links from node $i$ is given as $d_i := \sum_{j\in \partial i} w_{ij}$, the degree matrix $\bm{\mathcal{D}} := \mathrm{diag}(d_1,\,\dots\,d_n)$, 
where $\partial i$ denotes the set of out-neighbors of node $i$. 
Finally, Laplacian matrix $\bm{\mathcal{L}}$ is defined as $\bm{\mathcal{L}} := \bm{\mathcal{D}} - \bm{\mathcal{A}}$. 

Let ${}^t\!\bm{m} = (m_1,\,\dots,\,m_n)$ denote a left eigenvector associated with the eigenvalue $0$ of Laplacian matrix $\bm{\mathcal{L}}$. 
We say that the directed graph is {\em symmetrizable} if and only if $m_i>0$ for all $i\in V$ and 
\[
m_i \, w_{ij} = m_j \, w_{ji}, 
\]
for all pairs of adjacent nodes $(i \rightarrow j) \in E$. 
Hereafter, we denote the Laplacian matrix of a symmetrizable directed graph as $\bm{\mathcal{L}}_0$. 
The Laplacian matrix $\bm{\mathcal{L}}_0$ can be transformed into a symmetric matrix $\bm{S}_0$ by the similarity transformation 
using $\bm{M} := \mathrm{diag}(m_1,\,\dots,\,m_n)$, as 
\[
\bm{S}_0 := \bm{M}^{+1/2} \, \bm{\mathcal{L}}_0 \, \bm{M}^{-1/2}; 
\]
$\bm{S}_0$ and $\bm{\mathcal{L}}_0$ have the same eigenvalues. 
Furthermore, the eigenvalues are nonnegative and we sort them as 
\[
0 = \lambda_0 \le \lambda_1 \le \cdots \le \lambda_{n-1}. 
\]
In addition, we choose the eigenvectors $\bm{v}_\mu$ of $\bm{S}_0$ associated with $\lambda_\mu$ in such a way that the eigenvectors 
form an orthonormal eigenbasis, $\bm{v}_\mu\cdot\bm{v}_\nu=\delta_{\mu\nu}$. 

Next, we consider a simple and universal interaction model among users through recourse to the concept of the minimal model \cite{aida2018}.
We assume that the state of node $i$ at time $t$ (representing a user in the online social network) can be described by a one-dimensional parameter, $x_i(t)$. 
Also, it is assumed that the influence exists between adjacent nodes, such that they are influenced by the other's state quantities and tend to harmonize. 
We specifically assume that the strength of the influence between a pair of adjacent nodes, $i$ and $j$, is proportional to the absolute value of the difference between their state quantities, $|x_i (t)-x_j (t)|$.
Then, the equation of motion (EoM) of state vector $\bm{x}(t) = {}^t\!(x_1(t),\,\dots,\,x_n(t))$ can be denoted as 
\begin{align}
\frac{\dd^2}{\dd t^2} \, \bm{x}(t) = -\bm{\mathcal{L}} \, \bm{x}(t).
\label{eq:EoM}
\end{align}
This equation is called the {\em wave equation} on networks, and the above modeling is called the oscillation model on networks. 

We can calculate the oscillation energy of the whole network from the solution, $\bm{x}(t)$, of EoM~(\ref{eq:EoM}). 
In particular, if the social network is a symmetrizable directed graph, 
we can calculate the oscillation energy of each node from that of the whole network and, furthermore, the oscillation energy of each node gives a generalized notion of node centrality. 
Node centrality is a quantitative index indicating how important a particular node is in a network, and there are various different node centrality measures depending on the definition of importance used.
The representative indices are the degree centrality and the betweenness centrality, but the oscillation model on networks gives a framework that can explain both indices in a unified manner.
For example, if the weight of all links is $1$, the oscillation energy for each node becomes the degree centrality for the non-biased usage condition of the network.
Also, by taking the number of routes passing through a link (or the amount of passing traffic) as the link weight, the oscillation energy for each node gives an value related to the betweenness centrality, again for the non-biased usage condition of the network.
In particular, the oscillation energy of each node can generalize the node centrality even in various network usage situations 
such as having a biased usage condition where a specific node is the information source \cite{takano2018}.

If the social network is not symmetrizable, the oscillation energy of the whole network may diverge with time depending on the network structure.
This corresponds to the phenomenon where the strength of the user dynamics activity on the network diverges, like flaming on online social networks.
Since it is known that such divergence is not generated by symmetrizable directed graphs, 
the oscillation model on networks gives a model offers a generation mechanism of explosive user dynamics caused by the network structure \cite{takano2018}.

\section{Fundamental Equation of the Oscillation Model on Networks}
\label{sec:fundamental-eq}
This section briefly summarizes the fundamental equation of the oscillation model on networks according to \cite{aida2018}.  

We specifically consider the situation where the network is not symmetrizable, and can be decomposed into a symmetrizable and one-way link parts. 
We first discuss the difficulty in expressing the solution of the wave equation (\ref{eq:EoM_Lambda}) as a product of the solutions arising from the decomposition. 
We then show that, by reducing the second-order differential equation (\ref{eq:EoM_Lambda}) to a first-order equation, we can obtain a product-form solution that reflects the decomposition. 

We start with the decomposition of Laplacian matrix $\bm{\mathcal{L}}$ into the Laplacian matrix of symmetrizable directed graph, $\bm{\mathcal{L}}_0$, 
and that of a one-way link graph, $\bm{\mathcal{L}}_\textrm{I}$, as 
\begin{align}
\bm{\mathcal{L}} = \bm{\mathcal{L}}_0 + \bm{\mathcal{L}}_\textrm{I}, 
\label{eq:Decomp_L}
\end{align}
where the one-way link graph is a directed graph that has at most only one-way links between nodes. 
The decomposition (\ref{eq:Decomp_L}) is not unique and any directed graph can be decomposed as shown in (\ref{eq:Decomp_L}). 
Since the non-uniqueness of the decomposition (\ref{eq:Decomp_L}) leads to the selection of orthogonal bases in the state space through the choice of $\bm{S}_0$, 
we can choose a convenient decomposition that makes the Laplacian matrix of a one-way link graph $\bm{\mathcal{L}}_\textrm{I}$ simple.

The cause of the divergence in the oscillation energy is the influence of the one-way link graph, since the divergence of the oscillation energy is not inherent in symmetrizable directed graphs.
In order to directly express the influence of a one-way link graph on a symmetrizable directed graph, let us rewrite the EoM using the coordinate system obtained by converting $\bm{\mathcal{L}}_0$ into a diagonal matrix.
Let the orthonormal basis determined from $\bm{S}_0$ based on the decomposition (\ref{eq:Decomp_L}) be $\{\bm{v}_\mu\}_{0\le\mu\le n-1}$. 
By using the orthogonal matrix $\bm{P} := [\bm{v}_0,\,\bm{v}_1,\,\dots,\,\bm{v}_{n-1}]$, $\bm{\mathcal{L}}_0$ can be diagonalized as 
\[
\bm{\Lambda}_0 := {}^t\!\bm{P} \, \bm{S}_0 \, \bm{P} = {}^t\!\bm{P} \left( \bm{M}^{+1/2} \, \bm{\mathcal{L}}_0 \, \bm{M}^{-1/2} \right) \bm{P}, 
\]
where $\bm{\Lambda}_0 = \mathrm{diag}(\lambda_0,\,\dots,\,\lambda_{n-1})$. 
Let us define $\bm{\psi}(t) := {}^t\!\bm{P}\,\bm{M}^{+1/2} \, \bm{x}(t)$ and $\bm{\Lambda}_\mathrm{I} := {}^t\!\bm{P}\left( \bm{M}^{+1/2} \, \bm{\mathcal{L}}_\mathrm{I} \, \bm{M}^{-1/2}\right)\bm{P}$. 
Then, the EoM~(\ref{eq:EoM}) can be transformed into 
\begin{align}
\frac{\dd^2 \bm{\psi}(t)}{\dd t^2} &=  -\bm{\Lambda} \, \bm{\psi}(t) =  -(\bm{\Lambda}_0 + \bm{\Lambda}_\mathrm{I})\,\bm{\psi}(t) 
\label{eq:EoM_Lambda}
\end{align}
where $\bm{\Lambda} :=  \bm{\Lambda}_0 + \bm{\Lambda}_\mathrm{I}$. 

The solution of the wave equation~(\ref{eq:EoM_Lambda}) for $\bm{\Lambda}_\mathrm{I}=\bm{\mathrm{O}}$ (null matrix) is easily obtained. 
In order to explicitly describe the causal relation of the influence of $\bm{\Lambda}_\mathrm{I}$, 
it is preferable that the solution of~(\ref{eq:EoM_Lambda}) be cast in product-form; 
it consists of the solutions of the wave equations related to $\bm{\Lambda}_0$ and $\bm{\Lambda}_\mathrm{I}$. 
Unfortunately, attempting  the product-form solution of the wave equation~(\ref{eq:EoM_Lambda}) will not succeed. 
This is because the wave equation~(\ref{eq:EoM_Lambda}) is a second-order differential equation with respect to time, 
so the equation yields an extra cross term.  
Hence, to obtain a first-order differential equation with respect to time, we define the following matrix, 
\begin{align}
\bm{\Omega}^2 = \bm{\Lambda} = \bm{\Lambda}_0+\bm{\Lambda}_\mathrm{I}. 
\label{eq:Omega^2}
\end{align}
This means $\bm{\Omega}$ is the square root of matrix $\bm{\Lambda}$ and it is unique if we choose $\bm{\Omega}$ to be semi-positive definite.
If we define $\bm{\Omega}_0 := \bm{\Lambda}_0^{1/2}$, 
the square root matrix $\bm{\Omega}$ is decomposed as
\begin{align}
\bm{\Omega} = \bm{\Omega}_0 + \bm{\Omega}_\mathrm{I}. 
\label{eq:decomp-Omega}
\end{align}

By using the diagonal matrix $\bm{M}$ that symmetrizes $\bm{\mathcal{L}}_0$ into $\bm{S}_0$, 
the square root matrices $\bm{\mathcal{H}}_0$ of $\bm{\mathcal{L}}_0$, and $\bm{\mathcal{H}}$ of $\bm{\mathcal{L}}$ are  defined, respectively, as  
\begin{align}
\bm{\mathcal{H}}_0 &:= \bm{M}^{-1/2} \, (\bm{P} \, \bm{\Omega}_0 \, {}^t\!\bm{P}) \, \bm{M}^{+1/2}, 
\notag\\
\bm{\mathcal{H}} &:= \bm{M}^{-1/2} \, (\bm{P} \, \bm{\Omega} \, {}^t\!\bm{P}) \, \bm{M}^{+1/2}. 
\notag
\end{align}
Also, we define $\bm{\mathcal{H}}_\textrm{I}$ by using the decomposition 
\begin{align}
\bm{\mathcal{H}} &= \bm{\mathcal{H}}_0 + \bm{\mathcal{H}}_\textrm{I}. 
\label{eq:decomp-H}
\end{align}
Note that $\bm{\mathcal{H}}_\textrm{I}$ is not the square root of $\bm{\mathcal{L}}_\textrm{I}$. 

By using the square root matrix $\bm{\Omega}$ of $\bm{\Lambda}$, we introduce the following two different wave equations: 
\begin{align}
+\ii \, \frac{\dd \, \bm{\psi}^+(t)}{\dd t} = \bm{\Omega}\, \bm{\psi}^+(t), \quad -\ii \, \frac{\dd \, \bm{\psi}^-(t)}{\dd t} = \bm{\Omega}\, \bm{\psi}^-(t). 
\label{eq:fundamental}
\end{align}
The solutions of the wave equations~(\ref{eq:fundamental}) satisfy the following equation (double sign correspondence) as 
\begin{align}
\scalebox{0.95}{$\displaystyle 
\frac{\dd^2 \bm{\psi}^\pm(t)}{\dd t^2} = \mp\ii \bm{\Omega}\,\frac{\dd\,\bm{\psi}^\pm(t)}{\dd t} = -\bm{\Omega}^2\,\bm{\psi}^\pm(t) 
= -(\bm{\Lambda}_0+\bm{\Lambda}_\mathrm{I})\,\bm{\psi}^\pm(t). 
\notag
$}
\end{align}
This means that the solutions of the wave equations~(\ref{eq:fundamental}) solve the original wave equation~(\ref{eq:EoM_Lambda}). 

Conversely, let us confirm that the solution of the wave equation~(\ref{eq:EoM_Lambda}) does not necessarily solve (\ref{eq:fundamental}). 
For constants $c^+$ and $c^-$, let us consider a linear combination of the solutions of the two different equations~(\ref{eq:fundamental}), 
$c^+ \, \bm{\psi}^+(t) + c^- \, \bm{\psi}^-(t)$. 
The linear combination solves (\ref{eq:EoM_Lambda})
\begin{align}
\scalebox{0.95}{$\displaystyle 
\frac{\dd^2}{\dd t^2}\, (c^+ \, \bm{\psi}^+(t) + c^- \, \bm{\psi}^-(t)) = \ii \,\bm{\Omega}\,\frac{\dd}{\dd t}\, (-c^+ \, \bm{\psi}^+(t) + c^- \, \bm{\psi}^-(t)) 
$}
\notag\\
\scalebox{0.95}{$\displaystyle 
=-\,\bm{\Omega}^2 \,(c^+ \,\bm{\psi}^+(t) + c^- \, \bm{\psi}^-(t)) = -\,\bm{\Lambda} \,(c^+ \,\bm{\psi}^+(t) + c^- \, \bm{\psi}^-(t)).
$}
\notag
\end{align}
However, the linear combination satisfies neither of the equations in (\ref{eq:fundamental}). 
This problem is discussed later. 

Next, let us consider the possibility of the product-form solution of $\bm{\psi}^\pm(t)$ for the wave equation~(\ref{eq:fundamental}).
The goal here is to write solution $\bm{\psi}^\pm(t)$ in product-form, i.e. $\bm{\psi}^\pm(t) = \bm{\Psi}^\pm_0(t) \, \bm{\psi}^\pm_\mathrm{I}(t)$ by using  solutions of the two wave equations with respect to $\bm{\Omega}_0$ and $\bm{\Omega}_\mathrm{I}$.
By choosing the initial condition of $\bm{\Psi}^\pm_0(0) = \bm{I}$ ($n\times n$ unit matrix), that is, $\bm{\psi}^\pm(0) = \bm{\psi}^\pm_\textrm{I}(0)$, 
and by using the decomposition~(\ref{eq:decomp-Omega}), we introduce the following differential equations: 
\begin{align}
\pm\ii \,\frac{\dd}{\dd t} \, \bm{\psi}^\pm_0(t) &= \bm{\Omega}_0 \,  \bm{\psi}^\pm_0(t), 
\label{eq:psi_0&psi_I-1}\\
\pm\ii \,\frac{\dd}{\dd t} \, \bm{\psi}^\pm_\mathrm{I}(t) &= \big(\bm{\Psi}^\pm_0(-t) \, \bm{\Omega}_\mathrm{I} \, \bm{\Psi}^\pm_0(t)\big) \,  \bm{\psi}^\pm_\mathrm{I}(t), 
\label{eq:psi_0&psi_I-2}
\end{align}
where $\bm{\Psi}^\pm_0(t)$ is the diagonal matrix with diagonals ${}^t\!\bm{\psi}^\pm_0(t)=(\psi^\pm_0(0;t),\,\psi^\pm_0(1;t),\,\dots,\,\psi^\pm_0(n-1;t))$, that is,
\[
\bm{\Psi}^\pm_0(t) = 
\scalebox{0.9}{$\displaystyle
\begin{bmatrix} 
\psi^\pm_0(0;t) & 0 & \hdots  & 0 \\
0 & \psi^\pm_0(1;t) & \ddots & \vdots\\
\vdots & \ddots & \ddots & 0\\
0 & 0 & 0 & \psi^\pm_0(n-1;t)
\end{bmatrix}. 
$}
\] 
From the initial condition $\bm{\Psi}^\pm_0(0) = \bm{I}$, 
\begin{align}
\bm{\Psi}^\pm_0(-t) = \bm{\Psi}^\pm_0(t)^{-1} = \bm{\Psi}^\mp_0(t). 
\label{eq:Psi_0}
\end{align}
For ${}^t\bm{\psi}^\pm_\mathrm{I}(t)=(\psi^\pm_{\mathrm{I}}(0;t),\,\psi^\pm_{\mathrm{I}}(1;t),\,\dots,\,\psi^\pm_{\mathrm{I}}(n-1;t))$, 
the structure of the product-form solution is expressed as 
\[
\bm{\psi}^\pm(t) = \bm{\Psi}^\pm_0(t) \, \bm{\psi}^\pm_\mathrm{I}(t) = 
\scalebox{0.9}{$\displaystyle
\begin{pmatrix} 
\psi^\pm_0(0;t)\,\psi^\pm_{\mathrm{I}}(0;t)\\
\psi^\pm_0(1;t)\,\psi^\pm_{\mathrm{I}}(1;t) \\
\vdots \\
\psi^\pm_0(n-1;t)\,\psi^\pm_{\mathrm{I}}(n-1;t)
\end{pmatrix}. 
$}
\]

By substituting $\bm{\psi}^\pm(t) = \bm{\Psi}^\pm_0(t) \, \bm{\psi}^\pm_\mathrm{I}(t)$ into the wave equations~(\ref{eq:fundamental}), 
and using the differential equations~(\ref{eq:psi_0&psi_I-1}) and (\ref{eq:psi_0&psi_I-2}), and the relation (\ref{eq:Psi_0}), 
we obtain  
\begin{align}
\pm\ii \, \frac{\dd \, \bm{\psi}^\pm(t)}{\dd t} &= \pm\ii \, \frac{\dd}{\dd t} \, (\bm{\Psi}^\pm_0(t) \, \bm{\psi}^\pm_\mathrm{I}(t))
\notag\\
&=  \bm{\Omega}_0\,\bm{\Psi}^\pm_0(t) \, \bm{\psi}^\pm_\mathrm{I}(t) 
\notag\\
&\quad\quad {} 
+ \bm{\Psi}_0^\pm(t) \,  \big(\bm{\Psi}^\pm_0(-t) \, \bm{\Omega}_\mathrm{I} \, \bm{\Psi}^\pm_0(t)\big) \,\bm{\psi}^\pm_\mathrm{I}(t)
\notag\\
&= (\bm{\Omega}_0 + \bm{\Omega}_\mathrm{I})\, \bm{\psi}^\pm(t) 
\notag\\
&= \bm{\Omega}\, \bm{\psi}^\pm(t). 
\notag
\end{align}
This implies that the attempt to derive the product-form solution has succeeded. 

Summarizing the above, the solution of the wave equation~(\ref{eq:fundamental}) is also the solution of the original wave equation~(\ref{eq:EoM_Lambda}) 
and can be expressed as the product-form solution with respect to  $\bm{\psi}^\pm_0(t)$ and $\bm{\psi}^\pm_\mathrm{I}(t)$. 
Therefore, the causal relation of the influence of the one-way link graph can be explicitly described.  

From the above examination, the wave equations~(\ref{eq:fundamental}) that describe the causality of the oscillation dynamics can be 
considered as more fundamental than the original wave equation~(\ref{eq:EoM_Lambda}) (or the original EoM~(\ref{eq:EoM})). 
For this reason, we call the wave equations~(\ref{eq:fundamental}) the fundamental equations of oscillation dynamics on directed graphs.
Similarly, the first-order differential equations with respect to time that the original EoM~(\ref{eq:EoM}) can be rewritten into are
\begin{align}
+\ii \, \frac{\dd \, \bm{x}^+(t)}{\dd t} &= \bm{\mathcal{H}}\, \bm{x}^+(t), \quad -\ii \, \frac{\dd \, \bm{x}^-(t)}{\dd t} = \bm{\mathcal{H}}\, \bm{x}^-(t), 
\label{eq:fundamental-mathcalH}
\end{align}
they are the fundamental equations that are mathematically equivalent to (\ref{eq:fundamental}).

\section{Fundamental Equation and Quantum Theory}
\label{sec:quantum}
Let us rewrite the fundamental equations~(\ref{eq:fundamental-mathcalH}) into a single equation. 

First, we set the components of the $n$-dimensional vectors $\bm{x}^+(t)$ and $\bm{x}^-(t)$ as 
\begin{align}
\bm{x}^+(t) = {}^t\!(x^+_1(t),\,x^+_2(t),\,\dots,\,\psi^+_n(t)),
\notag\\
\bm{x}^-(t) = {}^t\!(x^-_1(t),\,x^-_2(t),\,\dots,\,\psi^-_n(t)).
\notag
\end{align}
By combining them, we define the new $2n$-dimensional vector $\bm{\hat{x}}(t)$ as 
\begin{align}
\bm{\hat{x}}(t) := {}^t\!(x^+_1(t),\,x^-_1(t),\,x^+_2(t),\,x^-_2(t),\,\dots,\,x^+_n(t),\,x^-_n(t)). 
\notag
\end{align}
Also, for Laplacian matrix $\bm{\mathcal{L}}$, we define the following $2n\times 2n$ square matrix 
\begin{align}
\bm{\mathcal{\hat{L}}} := \bm{\mathcal{L}} \otimes \bm{E}, 
\end{align}
where $\bm{E}$ denotes the $2\times 2$ unit matrix and $\otimes$ denotes Kronecker product \cite{brewer},  
which is, for  $\bm{\mathcal{L}} = [\mathcal{L}_{ij}]_{1\le i, j \le n}$, 
\begin{align}
\bm{\mathcal{\hat{L}}} &= 
\scalebox{0.8}{$\displaystyle
\begin{bmatrix}
\mathcal{L}_{11} \, \bm{E}  \,& \mathcal{L}_{12} \, \bm{E}  \,& \cdots & \mathcal{L}_{1n} \, \bm{E}\\
\mathcal{L}_{21} \, \bm{E}  \,& \mathcal{L}_{22} \, \bm{E}  \,& \cdots & \mathcal{L}_{2n} \, \bm{E}\\
\mathcal{L}_{31} \, \bm{E}  \,& \mathcal{L}_{32} \, \bm{E}  \,& \cdots & \mathcal{L}_{3n} \, \bm{E}\\
\vdots & \vdots & \ddots & \vdots \\
\mathcal{L}_{n1} \, \bm{E}  \,& \mathcal{L}_{n2} \, \bm{E}  \,& \cdots & \mathcal{L}_{nn} \, \bm{E}\\
\end{bmatrix}
$}
\notag\\
&=
\scalebox{0.7}{$\displaystyle
\begin{bmatrix}
\mathcal{L}_{11} & 0 & \mathcal{L}_{12} & 0 & \cdots & \mathcal{L}_{1n} & 0\\
0 & \mathcal{L}_{11} & 0 & \mathcal{L}_{12} & \cdots & 0 & \mathcal{L}_{1n} \\
\mathcal{L}_{21} & 0 & \mathcal{L}_{22} & 0 & \cdots & \mathcal{L}_{2n} & 0\\
0 & \mathcal{L}_{21} & 0 & \mathcal{L}_{22} & \cdots & 0 & \mathcal{L}_{2n} \\
\mathcal{L}_{31} & 0 & \mathcal{L}_{32} & 0 & \cdots & \mathcal{L}_{3n} & 0\\
0 & \mathcal{L}_{31} & 0 & \mathcal{L}_{32} & \cdots & 0 & \mathcal{L}_{3n} \\
\vdots & \vdots & \vdots & \vdots & \ddots & \vdots & \vdots\\
\mathcal{L}_{n-1,1} & 0 & \mathcal{L}_{n-1,2} & 0 & \cdots & \mathcal{L}_{n-1,n} & 0\\
0 & \mathcal{L}_{n-11} & 0 & \mathcal{L}_{n-1,2} & \cdots & 0 & \mathcal{L}_{n-1,n} \\
\mathcal{L}_{n1} & 0 & \mathcal{L}_{n2} & 0 & \cdots & \mathcal{L}_{nn} & 0\\
0 & \mathcal{L}_{n1} & 0 & \mathcal{L}_{n2} & \cdots & 0 & \mathcal{L}_{nn} \\
\end{bmatrix}.
$}
\notag
\end{align}
In order to express both $\bm{\mathcal{L}} = \bm{\mathcal{H}}^2$ and $\bm{\mathcal{L}} = (-\bm{\mathcal{H}})^2$ 
simultaneously, the square root $\bm{\mathcal{\hat{H}}}$ of $\bm{\mathcal{\hat{L}}}$ is defined as 
\begin{align}
\bm{\mathcal{\hat{H}}} &= \bm{\mathcal{H}} \otimes 
\scalebox{0.8}{$\displaystyle 
\begin{bmatrix}
1 & 0 \\
0 & -1
\end{bmatrix}. 
$}
\label{eq:2nx2n_H}
\end{align}
$\bm{\mathcal{\hat{H}}}$ satisfies $\bm{\mathcal{\hat{L}}} = \bm{\mathcal{\hat{H}}}^2$. 

By using the above $2n$-dimensional notations, the components of the fundamental equations~(\ref{eq:fundamental-mathcalH}) 
can be expressed as the one equation of 
\begin{align}
\ii \, \frac{\dd \, \bm{\hat{x}}(t)}{\dd t} &= \bm{\mathcal{\hat{H}}}\, \bm{\hat{x}}(t). 
\label{eq:fundamental-mathcalH2}
\end{align}
It is worth to note that this equation has essentially the same structure as the Dirac equation found in 
relativistic quantum theory~\cite{aida2018,Bjorken1965}. 

\section{Problems with the Fundamental Equation of the Oscillation Model}
The two expressions of the fundamental equations~(\ref{eq:fundamental}) and (\ref{eq:fundamental-mathcalH}) are mathematically 
equivalent and they can be transformed into each other by using a simple linear transformation. 
On the other hand, there are two crucial problems with the fundamental equation~(\ref{eq:fundamental-mathcalH}) as listed below. 
\begin{itemize}
\item The solutions of the fundamental equation~(\ref{eq:fundamental-mathcalH}) are also solutions of the original wave equation~(\ref{eq:EoM}). 
Unfortunately, the converse is not true, as shown in Sec.~\ref{sec:fundamental-eq}. 
If we are to claim that the fundamental equation~(\ref{eq:fundamental-mathcalH}) is really {\em fundamental}, it should be possible to derive all solutions of the original wave equation~(\ref{eq:EoM}) from the fundamental equation~(\ref{eq:fundamental-mathcalH}). 
\item The square root matrix $\bm{\mathcal{H}}$ of the Laplacian matrix that is appeared in the fundamental equation~(\ref{eq:fundamental-mathcalH}) 
does not reflect the structure of social networks described by the Laplacian matrix $\bm{\mathcal{L}}$. 
Since it is unacceptable in practice to hypothesize some direct relationships between nodes 
where links do not exist in the social network structure described by the Laplacian matrix, 
the Laplacian matrix $\bm{\mathcal{L}}$ and its square root matrix $\bm{\mathcal{H}}$ should have completely identical link structures.
\end{itemize}

\begin{figure}[t]
\begin{center}
\begin{tabular}{cc}
\includegraphics[width=0.41\linewidth]{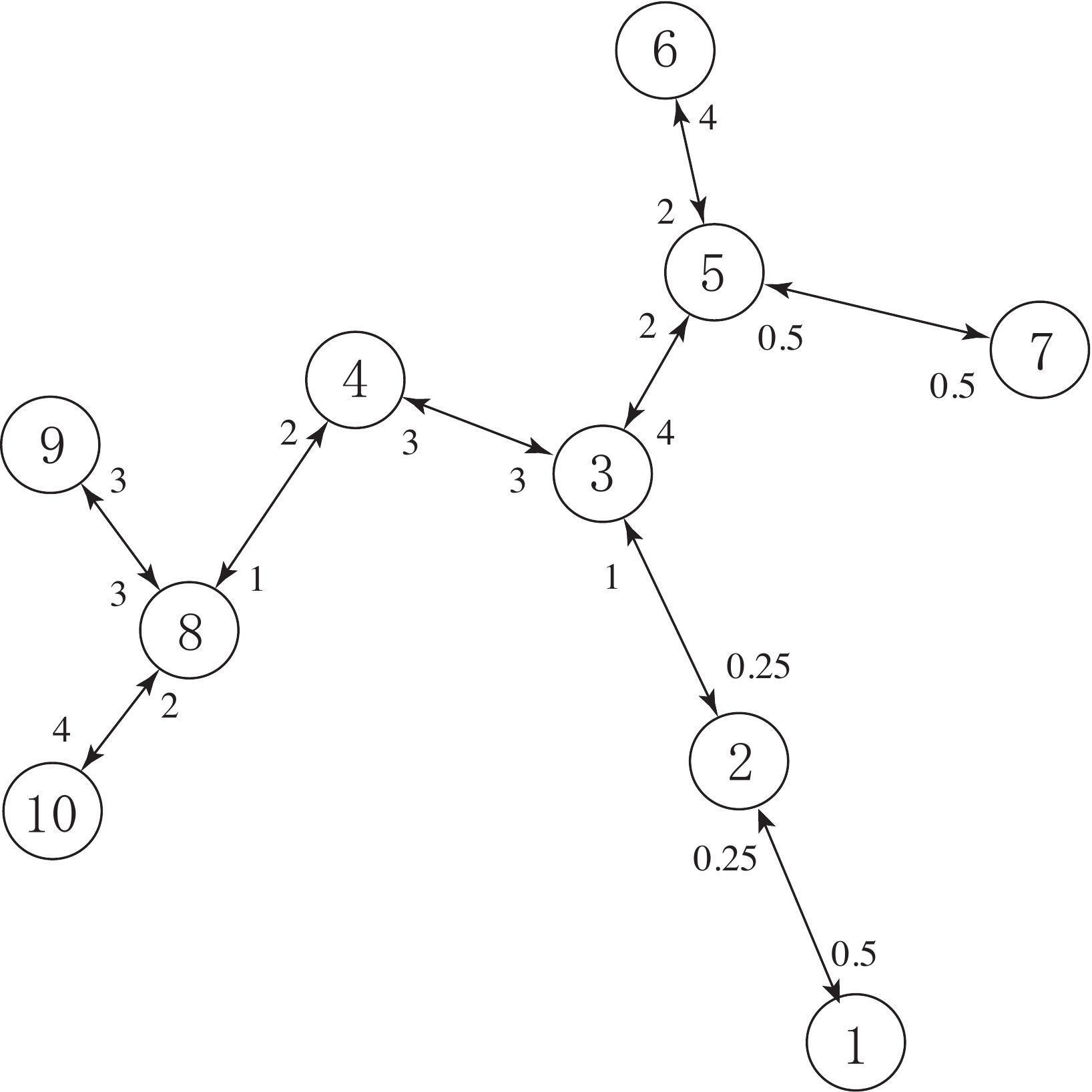} & 
\includegraphics[width=0.48\linewidth]{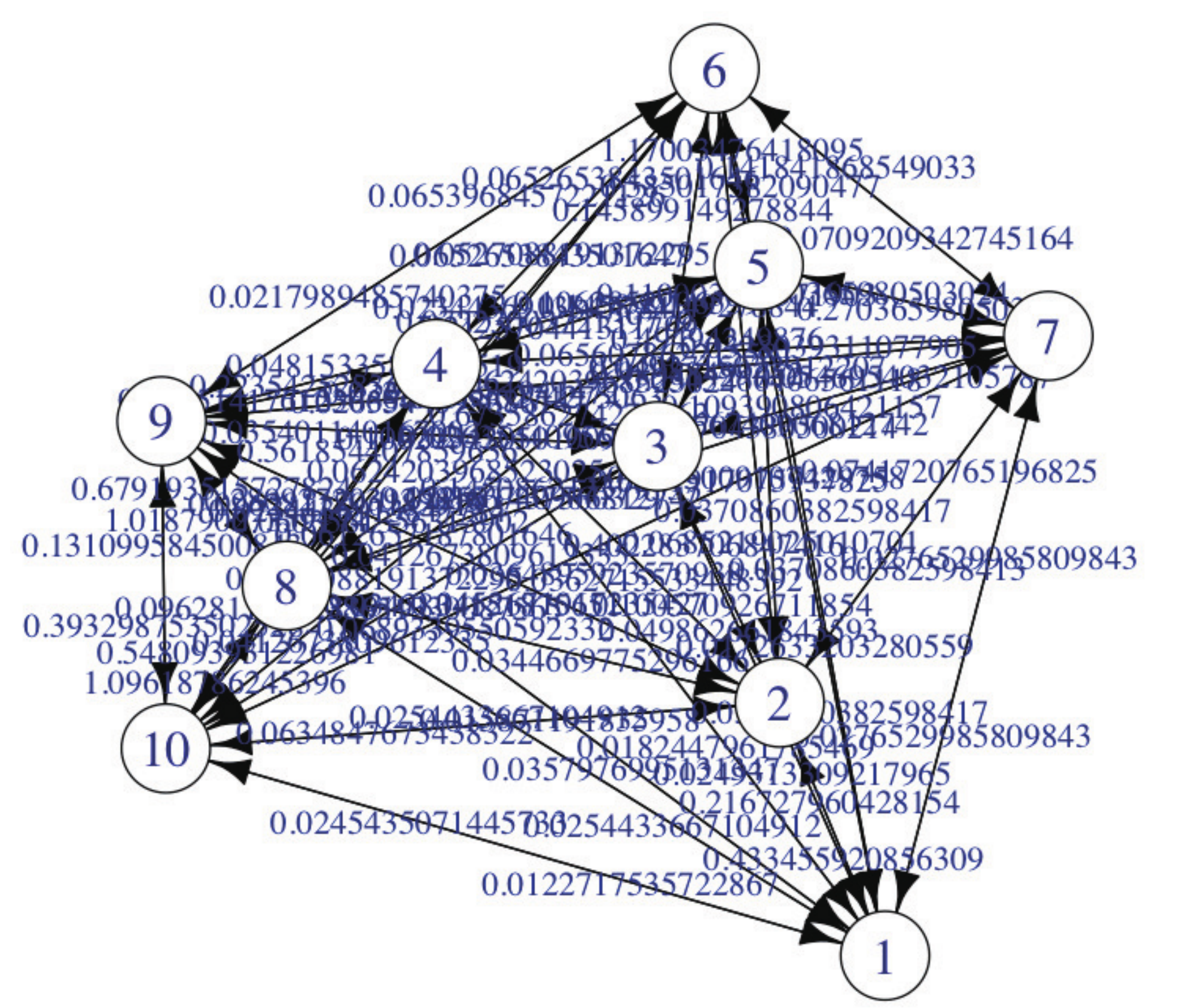}\\
link structure of $\bm{\mathcal{L}}$ & link structure of $\bm{\mathcal{H}}$ 
\end{tabular}
\caption{Link structures and link weights described by Laplacian matrix $\bm{\mathcal{L}}$ and its square root matrix $\bm{\mathcal{H}}$}
\label{fig:L&H}
\end{center}
\end{figure}

Here, we describe the latter problem via an example. 
The figure on the left of Fig.~\ref{fig:L&H} shows an example of a social network structure. 
For the Laplacian matrix $\bm{\mathcal{L}}$ describing the left figure, the figure on the right describes link structure of the square root matrix $\bm{\mathcal{H}}$ of $\bm{\mathcal{L}}$. 
Even if the structure of a social network is sparse, the link structure of its square root matrix is a complete graph, in general. 
This means that some direct relationships exist between all users yielding an unacceptable situation. 
Conversely, if we give $\bm{\mathcal{H}}$ as a sparse matrix, $\bm{\mathcal{L}}$ is also sparse. 
However, their link structures are not identical, in general, which is also an unacceptable situation.  

To solve these problems at the same time, we discuss the following proposition in the next section. 
\begin{theorem}
By utilizing one advantage of the structure of the $2n$-dimensional wave equation (\ref{eq:fundamental-mathcalH2}), the following two properties hold simultaneously: 
\begin{itemize}
\item The fundamental equation can generate all solutions of the original wave equation (\ref{eq:EoM}), and 
\item the matrix $\bm{\mathcal{\hat{H}}}$ appearing in the fundamental equation can be chosen so that  its link structure completely matches the link structure represented by the Laplacian matrix.
\end{itemize}
\end{theorem}

\section{Fundamental Equation Reflecting Social Network Structure}
\label{sec:nilpotent}
As shown in (\ref{eq:2nx2n_H}), the square root matrix $\bm{\mathcal{\hat{H}}}$ of $\bm{\mathcal{\hat{L}}} := \bm{\mathcal{L}}\otimes\bm{E}$ 
is chosen not as a semi-positive definite matrix, so the choice of $\bm{\mathcal{\hat{H}}}$ is not unique. 
The first attempt utilizes this degree of freedom of the choice
 to yield matching link structures, $\bm{\mathcal{\hat{H}}}$ and $\bm{\mathcal{\hat{L}}}$. 

First, we decompose $\bm{\mathcal{H}}$ into diagonal matrix $\bm{\mathcal{H}}^\mathrm{(d)}$ and 
the other matrix $-\bm{\mathcal{H}}^\mathrm{(a)}$, which has only non-diagonal components, as  
\[
\bm{\mathcal{H}} = \bm{\mathcal{H}}^\mathrm{(d)} - \bm{\mathcal{H}}^\mathrm{(a)}. 
\]
Since $\bm{\mathcal{H}}^2 = \bm{\mathcal{L}}$ and 
\begin{align}
\bm{\mathcal{H}}^2 &= ( \bm{\mathcal{H}}^\mathrm{(d)} - \bm{\mathcal{H}}^\mathrm{(a)})^2
\notag\\
&
= (\bm{\mathcal{H}}^\mathrm{(d)})^2 - \bm{\mathcal{H}}^\mathrm{(d)} \, \bm{\mathcal{H}}^\mathrm{(a)}
- \bm{\mathcal{H}}^\mathrm{(a)} \, \bm{\mathcal{H}}^\mathrm{(d)} + (\bm{\mathcal{H}}^\mathrm{(a)})^2, 
\notag
\end{align}
the link structures of both $\bm{\mathcal{H}}$ and $\bm{\mathcal{L}}$ are identical, if $(\bm{\mathcal{H}}^\mathrm{(a)})^2 = \bm{\mathrm{O}}$.   
In order to realize this relation, we consider the following $2n\times 2n$ matrix 
\begin{align}
\bm{\mathcal{\hat{H}}}^\mathrm{(a)} 
&=  \bm{\mathcal{H}}^\mathrm{(a)} \otimes \frac{1}{2}
\scalebox{0.8}{$\displaystyle 
\begin{bmatrix}
1 & 1\\
-1 & -1
\end{bmatrix}.
$}
\end{align}
The $2\times 2$ matrix used here exhibits nilpotency and so has the following property 
\[
\scalebox{0.8}{$\displaystyle 
\begin{bmatrix}
1 & 1\\
-1 & -1
\end{bmatrix}
$}^2 =
\scalebox{0.8}{$\displaystyle 
\begin{bmatrix}
0 & 0\\
0 & 0
\end{bmatrix},
$}
\]
 so $(\bm{\mathcal{\hat{H}}}^\mathrm{(a)})^2=\bm{\mathrm{O}}$. 

Let this nilpotent $2\times 2$ matrix be $\bm{X}$; by choosing a certain $2\times 2$ matrix $\bm{Y}$, we introduce 
\[
\bm{\mathcal{\hat{H}}} = \bm{\mathcal{H}}^\mathrm{(d)}\otimes \bm{Y}  - \bm{\mathcal{H}}^\mathrm{(a)} \otimes \bm{X}. 
\]
Here, we consider the possibility of whether or not the following relation is realized:
\[
\bm{\mathcal{\hat{H}}}^2 = \bm{\mathcal{L}}\otimes\bm{E}. 
\]
From the expansion of $\bm{\mathcal{\hat{H}}}^2$, we obtain
\begin{align}
\bm{\mathcal{\hat{H}}}^2 &= (\bm{\mathcal{H}}^\mathrm{(d)}\otimes \bm{Y}  - \bm{\mathcal{H}}^\mathrm{(a)} \otimes \bm{X})^2
\notag\\
&= (\bm{\mathcal{H}}^\mathrm{(d)})^2\otimes \bm{Y}^2  - (\bm{\mathcal{H}}^\mathrm{(d)}\,\bm{\mathcal{H}}^\mathrm{(a)}) \otimes (\bm{Y} \bm{X}) 
\notag\\
&\qquad {}- (\bm{\mathcal{H}}^\mathrm{(a)}\,\bm{\mathcal{H}}^\mathrm{(d)}) \otimes (\bm{X} \bm{Y}) + (\bm{\mathcal{H}}^\mathrm{(a)})^2 \otimes \bm{X}^2
\notag\\
&= (\bm{\mathcal{H}}^\mathrm{(d)})^2\otimes \bm{Y}^2  - (\bm{\mathcal{H}}^\mathrm{(d)}\,\bm{\mathcal{H}}^\mathrm{(a)}) \otimes (\bm{Y} \bm{X}) 
\notag\\
&\qquad {}
- (\bm{\mathcal{H}}^\mathrm{(a)}\,\bm{\mathcal{H}}^\mathrm{(d)}) \otimes (\bm{X} \bm{Y}). 
\notag
\end{align}
Therefore, the sufficient condition for $\bm{\mathcal{\hat{H}}}^2 = \bm{\mathcal{L}}\otimes\bm{E}$ can be written as 
\begin{align}
\bm{Y}^2 = \bm{E}, \quad \bm{X} \bm{Y} = \bm{Y} \bm{X} = \bm{E}, \quad \bm{X}^2 = \bm{\mathrm{O}}. 
\label{eq:Y}
\end{align}
Here, the second condition implies $\bm{Y} = \bm{X}^{-1}$, but it is known that no nilpotent matrix has an inverse.
Thus we cannot choose $\bm{Y}$ that satisfies the condition  (\ref{eq:Y}). 

The next step is to relax the condition $\bm{Y} = \bm{X}^{-1}$.  As one example that satisfies the following relation
\begin{align}
\bm{Y}^2 = \bm{E}, \quad \bm{X}^2 = \bm{\mathrm{O}}, 
\label{eq:Y2}
\end{align}
let us consider the following matrix~\cite{aida2020} 
\begin{align}
\bm{\mathcal{\hat{H}}} &:= \bm{\mathcal{\hat{H}}}^\mathrm{(d)} - \bm{\mathcal{\hat{H}}}^\mathrm{(a)} 
\notag\\
&
=  \bm{\mathcal{H}}^\mathrm{(d)} \otimes
\scalebox{0.8}{$\displaystyle 
\begin{bmatrix}
1 & 0\\
0 & -1
\end{bmatrix}
$}
- \bm{\mathcal{H}}^\mathrm{(a)} \otimes \frac{1}{2}
\scalebox{0.8}{$\displaystyle 
\begin{bmatrix}
1 & 1\\
-1 & -1
\end{bmatrix}. 
$}
\label{eq:H-nilpotent}
\end{align}
The corresponding fundamental equation is expressed as 
\begin{align}
\ii \, \frac{\dd \, \bm{\hat{x}}(t)}{\dd t} &= \bm{\mathcal{\hat{H}}}\, \bm{\hat{x}}(t) 
\notag\\
&= \left(\bm{\mathcal{H}}^\mathrm{(d)} \otimes
\scalebox{0.8}{$\displaystyle 
\begin{bmatrix}
1 & 0\\
0 & -1
\end{bmatrix}
$}
- \bm{\mathcal{H}}^\mathrm{(a)} \otimes \frac{1}{2}
\scalebox{0.8}{$\displaystyle 
\begin{bmatrix}
1 & 1\\
-1 & -1
\end{bmatrix}
$}
\right) \bm{\hat{x}}(t). 
\label{fundamenrtal-eq_std}
\end{align}
Here, we obtain 
\begin{align}
\bm{\mathcal{\hat{H}}}^2 
&= (\bm{\mathcal{H}}^\mathrm{(d)})^2\otimes 
\scalebox{0.8}{$\displaystyle 
\begin{bmatrix}
1 & 0\\
0 & 1
\end{bmatrix}
$}
- (\bm{\mathcal{H}}^\mathrm{(d)}\,\bm{\mathcal{H}}^\mathrm{(a)}) \otimes  \frac{1}{2}
\scalebox{0.8}{$\displaystyle 
\begin{bmatrix}
1 & 1\\
1 & 1
\end{bmatrix}
$}
\notag\\
&\quad\quad\quad{}
- (\bm{\mathcal{H}}^\mathrm{(a)}\,\bm{\mathcal{H}}^\mathrm{(d)}) \otimes  \frac{1}{2}
\scalebox{0.8}{$\displaystyle 
\begin{bmatrix}
1 & -1\\
-1 & 1
\end{bmatrix}
$}
\notag\\
&= (\bm{\mathcal{H}}^\mathrm{(d)})^2\otimes  \bm{E}
- (\bm{\mathcal{H}}^\mathrm{(d)}\,\bm{\mathcal{H}}^\mathrm{(a)} + \bm{\mathcal{H}}^\mathrm{(a)}\,\bm{\mathcal{H}}^\mathrm{(d)}) \otimes \frac{1}{2}\,\bm{E}
\notag\\
&\qquad\qquad
{}- (\bm{\mathcal{H}}^\mathrm{(d)}\,\bm{\mathcal{H}}^\mathrm{(a)}
- \bm{\mathcal{H}}^\mathrm{(a)}\,\bm{\mathcal{H}}^\mathrm{(d)}) \otimes \frac{1}{2}
\scalebox{0.8}{$\displaystyle 
\begin{bmatrix}
0 & 1\\
1 & 0
\end{bmatrix}. 
$}
\label{eq:Laplacian_2n_mix}
\end{align}
So, if $\bm{\mathcal{H}}^\mathrm{(d)}\,\bm{\mathcal{H}}^\mathrm{(a)} = \bm{\mathcal{H}}^\mathrm{(a)}\,\bm{\mathcal{H}}^\mathrm{(d)}$, 
that is, $\bm{\mathcal{H}}^\mathrm{(d)}$ and $\bm{\mathcal{H}}^\mathrm{(a)}$ are commutable (the order of product of 
matrices is commutable), we obtain the relation $\bm{\mathcal{\hat{H}}}^2 = \bm{\mathcal{L}}\otimes\bm{E}$. 
However, the commutation relation does not hold in general, and it is limited to the case 
when $\bm{\mathcal{H}}^\mathrm{(d)}$ is proportional to the unit matrix $\bm{I}$. 

From the above discussion, it can be seen that for a general Laplacian matrix 
$\bm{\mathcal{\hat{H}}}^2 = \bm{\mathcal{L}}\otimes\bm{E}$ cannot be satisfied while matching the links actually present.
Conversely, in order to match the link structures of $2n\times 2n$ matrices $\bm{\mathcal{\hat{H}}}$ and $\bm{\mathcal{\hat{L}}}$, 
we can recognize that two solutions of the fundamental equations (\ref{eq:fundamental-mathcalH}) should be mixed by the influence 
of the third term of the right-hand side of (\ref{eq:Laplacian_2n_mix}).
Expressing such a mixture of solutions yields the benefit of expressing the fundamental equation~(\ref{eq:fundamental-mathcalH2}) by using a $2n$-dimensional vector and a $2n\times 2n$ square matrix. 

Halting the attempt to realize $\bm{\mathcal{\hat{H}}}^2 = \bm{\mathcal{L}}\otimes\bm{E}$, 
We aim to reproduce the original equation of motion by pursuing the benefit of expressing the fundamental equation as a $2n$-dimensional vector.
Here, as discussed in Sec.~\ref{sec:fundamental-eq}, remember that the solutions of the wave equations~(\ref{eq:fundamental}) 
and that of the original equation of motion~(\ref{eq:EoM_Lambda}) are not the same.
The solutions of the wave equations~(\ref{eq:fundamental}) are always the solution of the original equation of motion~(\ref{eq:EoM_Lambda}), 
but the linear combination of the solutions of the two different wave equations~(\ref{eq:fundamental}) is also a solution of (\ref{eq:EoM_Lambda}).
Therefore, no problem is created if $\bm{\mathcal{\hat{H}}}^2$ mixes the solutions of the two fundamental equations~(\ref{eq:fundamental}) even if they have different signs; on the contrary, it is a desirable situation.

From the fundamental equation~(\ref{fundamenrtal-eq_std}), the second derivative of $\bm{\hat{x}}(t)$ is written as 
\begin{align}
\frac{\dd^2 \, \bm{\hat{x}}(t)}{\dd t^2} &= -\ii \, \frac{\dd}{\dd t} \, \bm{\mathcal{\hat{H}}}\, \bm{\hat{x}}(t) 
= - \bm{\mathcal{\hat{H}}}^2\, \bm{\hat{x}}(t) 
\notag\\
&= 
\scalebox{0.8}{$\displaystyle 
- \Bigg( (\bm{\mathcal{H}}^\mathrm{(d)})^2\otimes  
\begin{bmatrix}
1 & 0\\
0 & 1
\end{bmatrix}
- (\bm{\mathcal{H}}^\mathrm{(d)}\,\bm{\mathcal{H}}^\mathrm{(a)} + \bm{\mathcal{H}}^\mathrm{(a)}\,\bm{\mathcal{H}}^\mathrm{(d)}) \otimes \frac{1}{2}
\begin{bmatrix}
1 & 0\\
0 & 1
\end{bmatrix}
$}
\notag\\
&\qquad\qquad\qquad
\scalebox{0.8}{$\displaystyle 
{}- (\bm{\mathcal{H}}^\mathrm{(d)}\,\bm{\mathcal{H}}^\mathrm{(a)}
- \bm{\mathcal{H}}^\mathrm{(a)}\,\bm{\mathcal{H}}^\mathrm{(d)}) \otimes \frac{1}{2}
\begin{bmatrix}
0 & 1\\
1 & 0
\end{bmatrix}
\Bigg) \bm{\hat{x}}(t). 
$}
\end{align}
By extracting the differential equations for $\bm{x}^+(t)$ and $\bm{x}^-(t)$, we obtain 
\begin{align}
\scalebox{0.95}{$\displaystyle
\frac{\dd^2 \, \bm{x}^+(t)}{\dd t^2}
$} &
\scalebox{0.95}{$\displaystyle
= - \left( (\bm{\mathcal{H}}^\mathrm{(d)})^2
- \frac{1}{2}\,(\bm{\mathcal{H}}^\mathrm{(d)}\,\bm{\mathcal{H}}^\mathrm{(a)} + \bm{\mathcal{H}}^\mathrm{(a)}\,\bm{\mathcal{H}}^\mathrm{(d)})\right)\,  \bm{x}^+(t)
$}
\notag\\
&
\scalebox{0.95}{$\displaystyle
\qquad
{}-  \left(-\frac{1}{2}\,(\bm{\mathcal{H}}^\mathrm{(d)}\,\bm{\mathcal{H}}^\mathrm{(a)}
- \bm{\mathcal{H}}^\mathrm{(a)}\,\bm{\mathcal{H}}^\mathrm{(d)})\right)  \bm{x}^-(t),
$}
\label{eq:x^+}\\
\scalebox{0.95}{$\displaystyle
\frac{\dd^2 \, \bm{x}^-(t)}{\dd t^2}
$}
&
\scalebox{0.95}{$\displaystyle
= - \left( (\bm{\mathcal{H}}^\mathrm{(d)})^2
- \frac{1}{2}\,(\bm{\mathcal{H}}^\mathrm{(d)}\,\bm{\mathcal{H}}^\mathrm{(a)} + \bm{\mathcal{H}}^\mathrm{(a)}\,\bm{\mathcal{H}}^\mathrm{(d)})\right)\,  \bm{x}^-(t)
$}
\notag\\
&
\scalebox{0.95}{$\displaystyle
\qquad
{}- \left(-\frac{1}{2}\,(\bm{\mathcal{H}}^\mathrm{(d)}\,\bm{\mathcal{H}}^\mathrm{(a)}
- \bm{\mathcal{H}}^\mathrm{(a)}\,\bm{\mathcal{H}}^\mathrm{(d)})\right)  \bm{x}^+(t). 
$}
\label{eq:x^-}
\end{align}
Here, by adding both sides of the differential equations (\ref{eq:x^+}) and  (\ref{eq:x^-}), we obtain 
\begin{align}
&\frac{\dd^2}{\dd t^2} \, (\bm{x}^+(t)+\bm{x}^-(t)) 
\notag\\
&
= - \left( (\bm{\mathcal{H}}^\mathrm{(d)})^2 - \bm{\mathcal{H}}^\mathrm{(d)}\,\bm{\mathcal{H}}^\mathrm{(a)}\right)\,  (\bm{x}^+(t)+\bm{x}^-(t)).  
\label{eq:wave-eq_2}
\end{align}
This equation corresponds to the original equation of motion~(\ref{eq:EoM}).  
In addition, the solutions $\bm{x}^+(t)$ and $\bm{x}^-(t)$ of the fundamental equations satisfy, respectively, 
the same fundamental equation even if multiplied by a constant, so equation~(\ref{eq:wave-eq_2}) shows that 
the linear combination of the solutions of the fundamental equations~(\ref{fundamenrtal-eq_std}) is also 
the solution of the original equation of motion~(\ref{eq:EoM}).

From the above, by setting the fundamental equation to (\ref{fundamenrtal-eq_std}), 
it is possible to not only perfectly match the link structures between nodes represented by $\bm{\mathcal{\hat{H}}}$ 
and $\bm{\mathcal{L}}$, but also generate all solutions of the original equation of motion (\ref{eq:EoM}).

The correspondences of $\bm{\mathcal{H}}$ to the Laplacian matrix $\bm{\mathcal{L}}$, the adjacency matrix $\bm{\mathcal{A}}$, 
and the degree matrix $\bm{\mathcal{D}}$ are obtained as 
\begin{align}
\bm{\mathcal{L}} &= (\bm{\mathcal{H}}^\mathrm{(d)})^2 - \bm{\mathcal{H}}^\mathrm{(d)}\,\bm{\mathcal{H}}^\mathrm{(a)}, 
\notag\\
\bm{\mathcal{A}} &= \bm{\mathcal{H}}^\mathrm{(d)}\,\bm{\mathcal{H}}^\mathrm{(a)}, 
\\
\bm{\mathcal{D}} &= (\bm{\mathcal{H}}^\mathrm{(d)})^2. 
\notag
\end{align}
More specifically, we obtain the simple relations of 
\begin{align}
\bm{\mathcal{H}}^\mathrm{(d)} = \mathrm{diag}\left(\sqrt{d_1},\,\dots,\,\sqrt{d_n}\right), 
\label{eq:H^d}
\end{align}
and, since $\bm{\mathcal{H}}^\mathrm{(a)} = (\bm{\mathcal{H}}^\mathrm{(d)})^{-1}\,\bm{\mathcal{A}}$, 
$\bm{\mathcal{H}}^\mathrm{(a)} = [\mathcal{H}^\mathrm{(a)}_{ij}]_{1 \le i,j \le n}$ is obtained by 
\begin{align}
\mathcal{H}^\mathrm{(a)}_{ij} := \left\{
\begin{array}{cl}
w_{ij}/\sqrt{d_i}, \quad &(i \rightarrow j) \in E,\\
0, \quad &(i \rightarrow j) \not\in E. \\
\end{array}
\right.
\label{eq:H^a}
\end{align}
Incidentally, the existence of simple relations (\ref{eq:H^d}) and (\ref{eq:H^a}) is due to the selection of 
the nilpotent matrix 
$\scriptsize
\begin{bmatrix}
1 & 1\\
-1 & -1
\end{bmatrix}. 
$
If we choose the nilpotent matrix of 
$\scriptsize
\begin{bmatrix}
1 & -1\\
1 & -1
\end{bmatrix}, 
$
the adjacent matrix is obtained as $\bm{\mathcal{A}} = \bm{\mathcal{H}}^\mathrm{(a)}\,\bm{\mathcal{H}}^\mathrm{(d)}$, 
and the relations are complicated. 
This is because it complicates the use of the property of the Laplacian matrix that the row sum is zero.  

\begin{figure}[t]
\begin{center}
\begin{tabular}{ccc}
\includegraphics[width=0.41\linewidth]{Laplacian_nw.pdf} & 
\hspace{5mm} &
\includegraphics[width=0.41\linewidth]{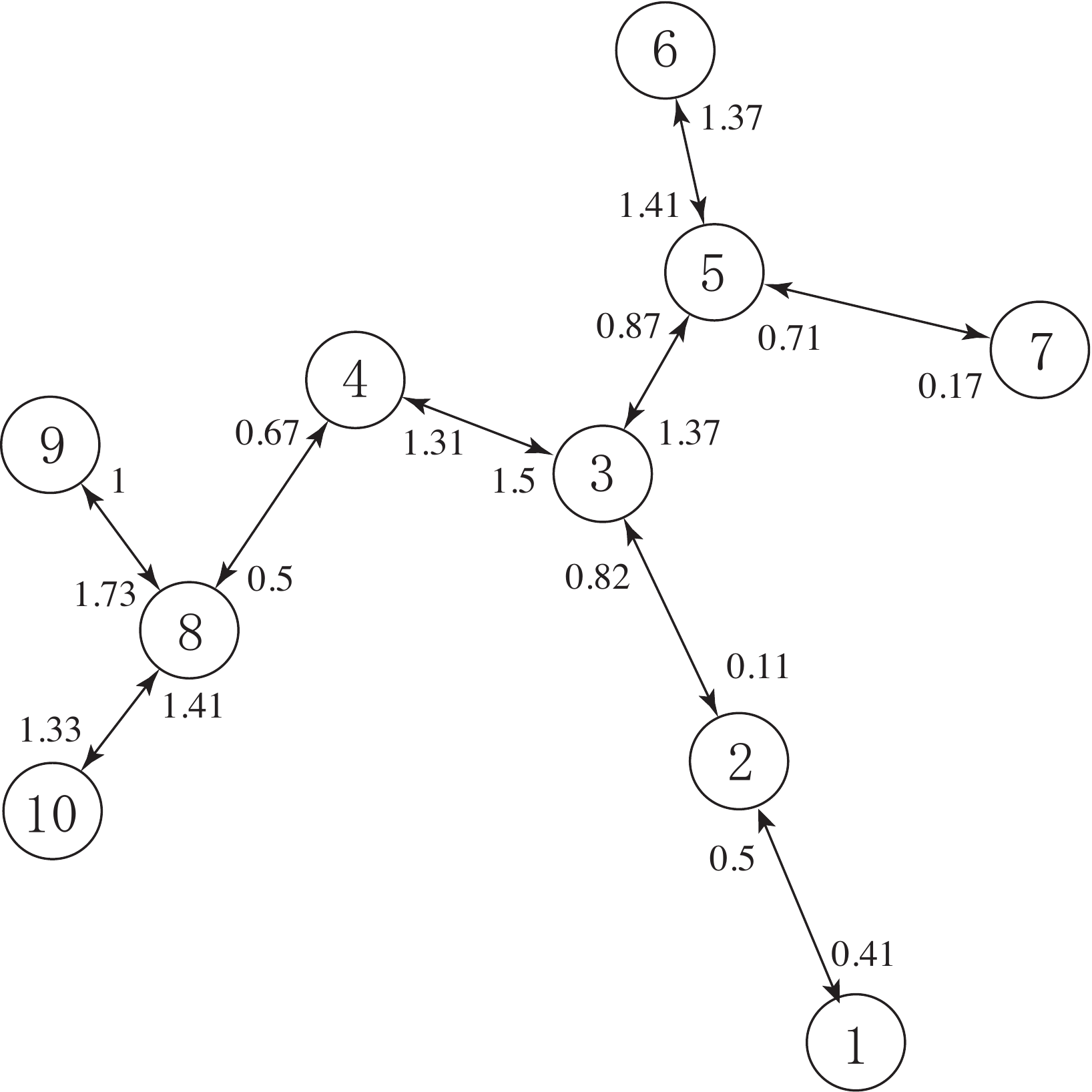}\\
link structure of $\bm{\mathcal{\hat{L}}}$ & & link structure of $\bm{\mathcal{\hat{H}}}$ 
\end{tabular}
\caption{Link structures and link weights described by the Laplacian matrix $\bm{\mathcal{\hat{L}}}$ and the proposed matrix $\bm{\mathcal{\hat{H}}}$}
\label{fig:hat_L&hat_H}
\end{center}
\end{figure}

Figure \ref{fig:hat_L&hat_H} shows an example of social network structures of $\bm{\mathcal{\hat{L}}}$ and $\bm{\mathcal{\hat{H}}}$. 

\section{Conclusions}
The two key remaining problems with the fundamental equation (\ref{eq:fundamental-mathcalH}) of the oscillation model on networks have been solved.
One is the problem that the solutions of (\ref{eq:fundamental-mathcalH}) (\ref{fundamenrtal-eq_std}) do not represent all the solutions of the original wave equation (\ref{eq:EoM}), and the other is that the link structures expressed by $\bm{\mathcal{H}}$ and those expressed by $\bm{\mathcal{L}}$ do not coincide.

This paper examined solutions to the latter problem and clarified that the two problems can be solved naturally and simultaneously.
The constraints of the matching the link structures of $\bm{\mathcal{H}}$ and $\bm{\mathcal{L}}$ 
while keeping the characteristic of modeling that clearly describes the causality of the fundamental equation (\ref{eq:fundamental}) is retained by considering the wave equation (\ref{eq:fundamental-mathcalH2}) as a $2n$-dimensional vector.
By utilizing one advantage of the structure of $2n$-dimensional wave equation (\ref{eq:fundamental-mathcalH2}), 
the solutions of the fundamental equations (\ref{eq:fundamental-mathcalH}) mix naturally and generate the solutions of the original equation of motion (\ref{eq:EoM}) of the $n$-dimensional vector, 
so solving the fundamental equation (\ref{fundamenrtal-eq_std}) gives all solutions of (\ref{eq:EoM}). 
While $\bm{\mathcal{\hat{H}}}^2 \not= \bm{\mathcal{\hat{L}}} := \bm{\mathcal{L}} \otimes \bm{E}$, 
$\bm{\mathcal{\hat{H}}}$ is not the square root matrix of $\bm{\mathcal{\hat{L}}}$, the following $n\times 2n$ matrix
\[
\bm{I} \otimes (1,1) = 
\scalebox{0.9}{$\displaystyle 
\begin{bmatrix}
1 & 1 & 0 & 0 & \cdots & 0 & 0\\
0 & 0 & 1 & 1 & \cdots & 0 & 0\\
\vdots &\vdots &\vdots &\vdots &\ddots &\vdots &\vdots \\
0 & 0 & 0 & 0 & \cdots & 1 & 1
\end{bmatrix}, 
$}
\]
can be used to obtain 
\begin{align}
(\bm{I} \otimes (1,1) ) \, \bm{\mathcal{\hat{H}}}^2 \,  \bm{\hat{x}} = \bm{\mathcal{L}} \,  \bm{x}. 
\end{align}
where $(\bm{I} \otimes (1,1) ) \,  \bm{\hat{x}} =  \bm{x}$. 

\section*{Acknowledgment}
This research was supported by Grant-in-Aid for Scientific Research (B) No.~17H01737 (2017--2019) and 
No.~19H04096 (2019--2021), and Grant-in-Aid for Scientific Research (C) No.~18K11271 (2018--2020) 
and No.~18K13777 (2018-2020) from the Japan Society for the Promotion of Science (JSPS).

%
%

\end{document}